\documentclass[twocolumn,tighten]{aastex631}

\usepackage{amsmath}
\usepackage{xspace}
\usepackage{multirow}
\usepackage{fancyapj}
\usepackage{mathtools}
\usepackage{xcolor}

\makeatletter
\def\restartappendixnumbering{\global\applettertrue
\setcounter{table}{0}
\setcounter{figure}{0}
\setcounter{equation}{0}
\def\thetable{\thesection\the\c@table}%
\renewcommand{\theHtable}{Supplement.\thetable}
\def\fnum@table{{\bf\tablename~\thetable}}%
\def\thefigure{\thesection\the\c@figure}%
\def\fnum@figure{{\bf\figurename~\thefigure}}%
}%
\makeatother

  {\list{}{\leftmargin=0.1in\rightmargin=0.1in}\item[]}%
  {\endlist}

\expandafter\def\csname editcolor1\endcsname{magenta}
\expandafter\def\csname editcolor2\endcsname{red}

\newcommand{\E}[1]{\ensuremath{\times 10^{#1}} }

\newcommand{\msol}{\ensuremath{M_{\odot}}\xspace}

\newcommand{\kev}{\rm\,keV\xspace}
\newcommand{\mev}{\rm\,MeV\xspace}
\newcommand{\hz}{\rm\,Hz\xspace}
\newcommand{\ks}{\rm\,ks\xspace}
\newcommand{\km}{\rm\,km\xspace}

\newcommand{\kpc}{\rm\,kpc\xspace}
\newcommand{\s}{\rm\,s\xspace}

\newcommand{\hr}{\rm\,hr\xspace}
\newcommand{\per}[1]{\rm\,#1\ensuremath{^{-1}}\xspace}
\newcommand{\persq}[1]{\rm\,#1\ensuremath{^{-2}}\xspace}
\newcommand{\lumcgs}{\rm\,erg\per{s}\xspace}
\newcommand{\fluxcgs}{{\rm\,erg{\per{s}}{\persq{cm}}\xspace}}

\newcommand{\porb}{\ensuremath{P_{\rm orb}}\xspace}
\newcommand{\asini}{\ensuremath{a_x \sin i}\xspace}
\newcommand{\tasc}{\ensuremath{T_{\rm asc}}\xspace}

\newcommand{\nustar}{\textrm{NuSTAR}\xspace}
\newcommand{\nicer}{\textrm{NICER}\xspace}

\newcommand{\src}{MAXI~J1816\xspace}
\newcommand{\srcfull}{MAXI~J1816$-$195\xspace}

\begin{document}

\title{The discovery of the 528.6 Hz accreting millisecond X-ray pulsar MAXI J1816--195}

\author[0000-0002-7252-0991]{Peter Bult}
\affiliation{Department of Astronomy, University of Maryland, College Park, MD 20742, USA}
\affiliation{Astrophysics Science Division, NASA Goddard Space Flight Center, Greenbelt, MD 20771, USA}

\author[0000-0002-3422-0074]{Diego Altamirano}
\affiliation{Physics \& Astronomy, University of Southampton, Southampton, Hampshire SO17 1BJ, UK}

\author{Zaven Arzoumanian}
\affiliation{Astrophysics Science Division, NASA Goddard Space Flight Center, Greenbelt, MD 20771, USA}

\author[0000-0001-8804-8946]{Deepto Chakrabarty}
\affiliation{MIT Kavli Institute for Astrophysics and Space Research, Massachusetts Institute of Technology, Cambridge, MA 02139, USA}

\author[0000-0002-4397-8370]{J\'er\^ome Chenevez}
\affiliation{DTU Space, Technical University of Denmark, Elektrovej 327-328, 2800 Kongens Lynby, Denmark}

\author[0000-0001-7828-7708]{Elizabeth C. Ferrara}
\affiliation{Department of Astronomy, University of Maryland, College Park, MD 20742, USA}
\affiliation{Astrophysics Science Division, NASA Goddard Space Flight Center, Greenbelt, MD 20771, USA}

\author[0000-0001-7115-2819]{Keith C. Gendreau}
\affiliation{Astrophysics Science Division, NASA Goddard Space Flight Center, Greenbelt, MD 20771, USA}

\author[0000-0002-6449-106X]{Sebastien~Guillot}
\affil{Institut de Recherche en Astrophysique et Plan\'{e}tologie, UPS-OMP, CNRS, CNES, 9 avenue du Colonel Roche, BP 44346, F-31028 Toulouse Cedex 4, France}

\author[0000-0002-3531-9842]{Tolga G\"uver}
\affiliation{Istanbul University, Science Faculty, Department of Astronomy and Space Sciences, Beyaz\i t, 34119, Istanbul, Turkey}
\affiliation{Istanbul University Observatory Research and Application Center, Istanbul University 34119, Istanbul Turkey}

\author[0000-0002-0207-9010]{Wataru Iwakiri}
\affiliation{Department of Physics, Faculty of Science and Engineering, Chuo University, 1-13-27 Kasuga, Bunkyo-ku, Tokyo 112-8551, Japan}

\author[0000-0002-6789-2723]{Gaurava K. Jaisawal} 
\affiliation{DTU Space, Technical University of Denmark, Elektrovej 327-328, 2800 Kongens Lynby, Denmark}

\author[0000-0001-9822-6937]{Giulio C. Mancuso}
\affiliation{Instituto Argentino de Radioastronom\'{\i}a (CCT-La Plata, CONICET; CICPBA), C.C. No. 5, 1894 Villa Elisa, Argentina}
\affiliation{Facultad de Ciencias Astron\'omicas y Geof\'{\i}sicas, Universidad Nacional de La Plata, Paseo del Bosque s/n, 1900 La Plata, Argentina}

\author[0000-0002-0380-0041]{Christian Malacaria}
\affiliation{International Space Science Institute (ISSI), Hallerstrasse 6, 3012 Bern, Switzerland}

\author[0000-0002-0940-6563]{Mason Ng}
\affiliation{MIT Kavli Institute for Astrophysics and Space Research, Massachusetts Institute of Technology, Cambridge, MA 02139, USA}

\author[0000-0002-0118-2649]{Andrea Sanna}
\affiliation{Dipartimento di Fisica, Universit\`a degli Studi di Cagliari, SP Monserrato-Sestu km 0.7, 09042 Monserrato, Italy}

\author[0000-0001-7681-5845]{Tod E. Strohmayer}
\affil{Astrophysics Science Division and Joint Space-Science Institute, NASA's Goddard Space Flight Center, Greenbelt, MD 20771, USA}

\author[0000-0002-9249-0515]{Zorawar Wadiasingh}
\affiliation{Department of Astronomy, University of Maryland, College Park, MD 20742, USA}
\affiliation{Astrophysics Science Division, NASA Goddard Space Flight Center, Greenbelt, MD 20771, USA}
\affiliation{Center for Research and Exploration in Space Science and Technology, NASA/GSFC, Greenbelt, Maryland 20771, USA}

\author[0000-0002-4013-5650]{Michael T. Wolff}
\affiliation{Space Science Division, U.S. Naval Research Laboratory, Washington, DC 20375-5352, USA}

\begin{abstract}\nolinenumbers
  We present the discovery of $528.6$ Hz pulsations in the new X-ray transient
  MAXI J1816--195. Using NICER, we observed the first recorded transient
  outburst from the neutron star low-mass X-ray binary MAXI J1816--195 over
  a period of 28 days. From a timing analysis of the $528.6$ Hz pulsations, we
  find that the binary system is well described as a circular orbit with an
  orbital period of 4.8 hours and a projected semi-major axis of $0.26$
  light-seconds for the pulsar, which constrains the mass of the donor star to
  $0.10-0.55\msol$. Additionally, we observed 15 thermonuclear X-ray bursts
  showing a gradual evolution in morphology over time, and a recurrence time as
  short as 1.4 hours. We did not detect evidence for photospheric radius
  expansion, placing an upper limit on the source distance of $8.6$ kpc.
\end{abstract}


\section{Introduction}
\label{sec:intro}
%\nolinenumbers

Accreting millisecond X-ray pulsars (AMXPs) are rapidly rotating neutron stars
that accrete matter from a binary companion \citep[see][for
reviews]{Patruno2021, DiSalvo2022}. The characteristic that sets these systems
apart from the wider population of low-mass X-ray binary systems is that they
exhibit a coherent pulsation which directly tracks the millisecond stellar rotation period
of the neutron star. Such pulsations are a useful diagnostic for
the accreting neutron star. For instance, the precise waveform of the
pulsations encodes information about the shape of the surface emission region and
the neutron star compactness, and thus its
equation of state \citep{Poutanen2003}; while tracking of the pulse arrival
times allows for a high precision measurement of the neutron star spin and
binary ephemeris, and may in principle be used to investigate the
torques acting on these millisecond pulsars (\citealt{Bildsten1998b, Psaltis1999b}; 
see \citet{Burderi2006, Hartman2008, Patruno2021, DiSalvo2022} for recent
discussions of different torque mechanisms that may play a role).

An enduring challenge to the study of AMXPs is that they are relatively rare.
Since the discovery of pulsations from SAX J$1808.4-3658$ in 1998
\citep{Wijnands1998b}, the population of AMXPs has grown at an average rate of
under one per year. All the known AMXPs are X-ray transients, and the accretion-powered
pulsations are only visible while the source is undergoing an X-ray outburst.
These outbursts typically last only a few days to a few weeks, and are
interspersed by several years to decades of inactivity \citep{Lasota2001, Hameury2020}. As such, the search for
new accreting millisecond pulsars remains an important task. 

In this letter we present the discovery of $528.6\hz$ pulsations from \srcfull
(henceforth \src). This system was first discovered as a new X-ray transient
with the MAXI Gas Slit Camera on 2022 June 7 \citep{AtelNegoro22a}, and an
initial source localization was provided by Swift shortly thereafter
\citep{AtelKennea22a}. Subsequent follow-up observations with the Neutron Star
Interior Composition Explorer (NICER) on 2022 June 8 revealed the presence of
$528.6\hz$ pulsations \citep{AtelBult22a}, identifying \src as an accreting
millisecond pulsar. 

Further monitoring with NICER revealed the 4.8\hr binary orbit of \src
\citep{AtelBult22b}. Additional Swift observations yielded an improved X-ray
localization \citep{AtelKennea22b}, after which likely counterparts to the
X-ray source were identified in infrared \citep{AtelKennea22b}, optical
\citep{AtelMartino22}, and radio \citep{AtelBeauchamp22, AtelBright22}.

In the following we present the detailed analysis underpinning the NICER
discovery of pulsations from \src, and the subsequent characterization of its
orbit. We describe the NICER monitoring campaign and provide a pulsar
timing analysis. Further, we present an analysis of fifteen thermonuclear X-ray bursts
observed from this system, and interpret the various implications from both the
bursts and pulsations on the nature of \src.  An analysis of the
(non-burst) X-ray spectroscopy will be presented elsewhere. 

\section{Observations}
We have monitored the 2022 outburst of \src extensively with NICER. Our
observations began on 2022 June 7 and continued through 2022 July 5, at
which time the source intensity had decreased to the background level.
These data are available under the NICER ObsIDs starting with 520282 and 553301.

The source coordinates used for instrument pointing evolved slightly over the
course of the monitoring campaign. Our first two observations were collected by
pointing at the NICER raster scan coordinates reported by \citet{AtelBult22a}.
The subsequent monitoring program used the initial Swift localization of
\citet{AtelKennea22a} until 2022 June 30, when we switched to the corrected
coordinates of \citet{AtelKennea22b}. These coordinates are all $<50\arcsec$
apart, which is much smaller than the offset angle at which the response of
NICER is significantly affected\footnote{
  \url{https://heasarc.gsfc.nasa.gov/docs/nicer/analysis_threads/cal-recommend/}
}. 

The data were processed using \textsc{nicerdas} version $9$ as distributed with
\textsc{heasoft} version $6.30$. To account for the small variations in the
pointing coordinates, we filter the data such that the angular offset relative to the pointing
coordinates is $<54\arcsec$\footnote{The default behavior of the NICER pipeline
is to calculate the pointing offset relative to the source coordinates.}.
Otherwise, we used standard filtering criteria; we included only
epochs which had an elevation angle $>15\arcdeg$, a bright Earth limb angle
$>30\arcdeg$, and were outside the South Atlantic Anomaly (SAA). 

After applying these filtering criteria, we were left with 92\ks clean
exposure. We corrected the clean data to the solar system barycenter using the
DE-430 planetary ephemeris \citep{Folkner2014} and source
coordinates of \citet{AtelKennea22b}.

Inspecting a $1\s$ time resolution light curve of both clean and unfiltered
data, we readily identify 15 thermonuclear (type I) X-ray bursts. Ten of these were observed in
full in the clean data, while another three were observed during SAA passages.
The remaining two bursts were truncated by the boundaries of the observations.

\section{Results}

\begin{figure}[t]
  \includegraphics[width=\linewidth]{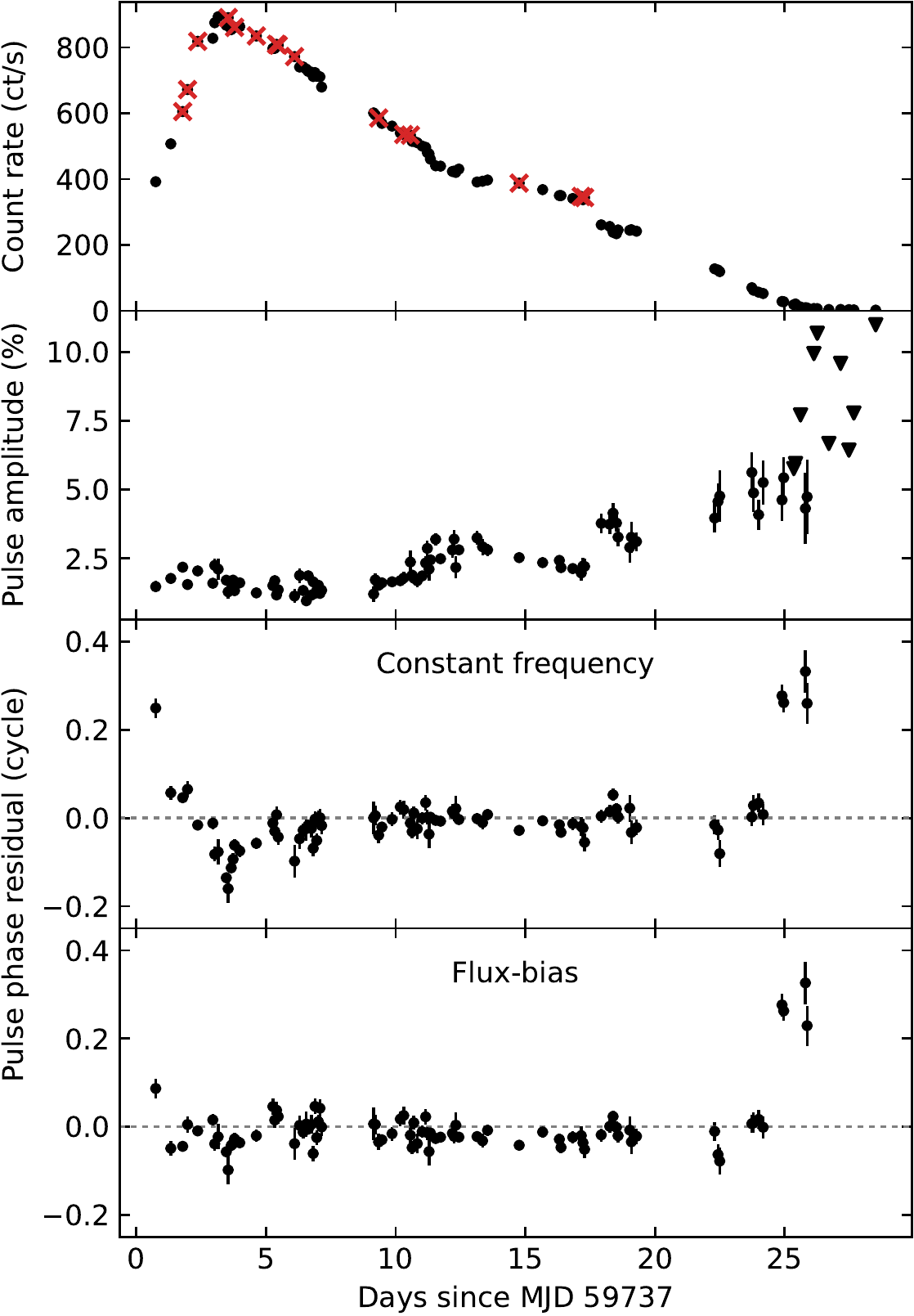}
  \caption{%
    Outburst evolution of \src relative to 2022 June 7 (MJD 59737). The top panel shows the $0.5-10\kev$ count-rate
    as a function of time, with each point representing a single \nicer
    pointing.  The X-ray bursts have been removed from this data, but their
    onset times are marked with red crosses. The second panel shows the
    fractional pulse amplitude in the $0.5-10\kev$ range using triangles to
    indicate 95\% upper limits for non-detections. Finally, the two bottom
    panels give the phase residuals relative to a constant frequency and
    flux-bias adjusted timing model (see Table \ref{tab:timing
    solution}).
  } 
  \label{fig:outburst}
\end{figure}

We began our analysis by considering the evolution of the outburst. We divided
the data into segments of continuous pointing, finding 93 such
pointings across the 28 days of monitoring with an exposure per pointing between
$150-2250\s$. For each pointing we calculated the average
count-rate in the $0.5-10\kev$ band, which we show in Figure
\ref{fig:outburst}. 

\subsection{Coherent timing}

We initially searched for the presence of pulsations by dividing the first
observation into $64\s$ segments and taking the Fourier transform of each
segment. Converting these transforms to an averaged power spectrum revealed
a high significance ($>6\sigma$) pulse signal at $528.6\hz$.

To investigate the pulse signal in greater detail, we determined the pulse
frequency that optimized the $Z^2_1$ score \citep{Buccheri1983} for each
pointing separately. The resulting pulse frequencies were found to show a clear
$4.8\hr$ periodic oscillation in time, revealing the orbit of the binary
system.  We fitted these frequency measurements using a sinusoid to extract an
initial estimate of the orbital period, $\porb$, the neutron star's projected semi-major axis,
$\asini$, and the time of its passage through the ascending node, $\tasc$.

Based on the initial timing solution, we corrected the photon arrival
times for the Doppler delays of the binary motion and then folded each
pointing to a pulse profile. We fitted these profiles using a constant plus two
harmonically related sinusoids, one fixed at the pulse frequency and the other
at twice that frequency. Either harmonic was deemed significant if its measured
amplitude was greater than three times the uncertainty, $A/\sigma_A > 3$. We
measured a significant amplitude for the fundamental pulsation in 83 out of 92 pointings, with
the non-detections all confined to the final days of the outburst where the
count-rate dropped below the background level. The second harmonic was only
required in two pointings near the peak of the outburst. We proceeded by
selecting the measured phases of the fundamental and converted them
to pulse arrival times. These arrival times were then
modeled with \textsc{tempo2} \citep{Hobbs2006}, using a constant pulse
frequency and a circular orbital model. We repeated this process of folding the
data and fitting the pulse arrival times until the timing solution converged. 

The parameters of the obtained timing solution are listed in Table
\ref{tab:timing solution}, while the resulting pulse amplitudes and phase
residuals are shown in Figure \ref{fig:outburst} (second and third panel). From
the figure it is clear that the timing solution describes the decay of the
outburst well, but leaves systematic residuals before $t=5$\,d and after
$t=25$\,d.

Including a frequency derivative in the model did not meaningfully improve the
quality of the fit. Instead, we found that we needed to include terms up to the fifth
frequency derivative before the structural phase residuals were reduced.
However, such a high-degree polynomial frequency model is plainly unphysical.

In an alternative approach, we attempted to fit the data using the flux-bias
model of \citet{Bult2020a}. We expressed the phase model as
\begin{equation}
  \label{eq:phase model}
  \varphi(t, F) = \varphi_0 + \nu_0 t + \varphi_{\rm orb}(t) + b F^\Gamma,
\end{equation}
where $\varphi_0$ is an arbitrary phase offset, $\nu_0$ the constant pulsar
spin frequency, and $\varphi_{\rm orb}(t)$ the phase correction associated with
the binary orbit. The final term on the right hand side adds a flux dependent
phase bias with scale factor $b$, and power law index $\Gamma$. We adopted the
$0.5-10\kev$ count-rate as a proportional substitute for flux and set the power law
index to $-1/5$ to model the effect of a phase drift imposed by the moving
magnetospheric boundary \citep{Bult2020a}.  This approach again only marginally
improved the fit. Leaving the power law index free to vary gives the model
sufficient flexibility to account for phase residuals during the outburst rise,
but leaves a discrete $0.3$ cycle jump in phase late in the outburst. We list the
parameters of this model in Table \ref{tab:timing solution} and show the phase residuals
in Figure \ref{fig:outburst} (fourth panel).

Finally, we investigated the energy dependence of the pulse waveform. We
divided the $0.5-10\kev$ energy range into 50 bins, such that each bin contains
a roughly equal number of photons. For each energy bin we folded the entire
dataset to a pulse profile using the flux-bias timing model and measured the
amplitude and phase of the fundamental pulsation. We found that the pulse phase
remained constant across all energy bins within measurement uncertainties,
while the fractional amplitude increases linearly with energy from
$(0.5\pm0.2)\%$ at $0.5\kev$ to $(6.4\pm0.1)\%$ at $10\kev$.

\begin{table}[t]
  %\centering
  \movetableright=-0.30in
  \caption{%
    Timing solution of \src
    \label{tab:timing solution}
  }
  \begin{tabular}{l l l h l}
    \hline \hline
    Parameter  &  Value  &  Uncertainty   \\ 
    \tableline
    Epoch (MJD)      & 59750             & ~         \\
    \porb (d)        & 0.20141878        & 5\E{-8}   \\
    \asini (lt-s)    & 0.262949          & 1.4\E{-5} \\
    \tasc (MJD)      & 59738.8756284     & 2.7\E{-6} \\
    Eccentricity     & $< 2\E{-4}$       & ~         \\
    \tableline
    \multicolumn{3}{l}{Constant frequency model}     \\
    \tableline
    $\nu_0$ (Hz)     &  528.611105819     & 1.0\E{-8} \\ 
    $\chi^2$ / dof   &  2167 / 78         & ~         \\
    \tableline
    \multicolumn{3}{l}{Flux-bias model}              \\
    \tableline
    $\nu_0$ (Hz)     &  528.611105950    & 1.4\E{-8} \\
    $b$ (cycles/rate$^\Gamma$)  &  4.7\E{-5}        & 5\E{-6}   \\
    $\Gamma$         &  1.2              & 0.2       \\
    $\chi^2$ / dof   &  213 / 72         & ~         \\
    \tableline
  \end{tabular}
  \flushleft
  \tablecomments{The orbital parameters listed are obtained by fitting the
  constant frequency model to the $t=5-25$\,d data. The orbital parameters
  obtained with the flux-bias model are consistent with listed values within
  quoted errors. All reported MJD are barycentric (TDB). Uncertainties are
  $1\sigma$ errors, upper limits are quoted at 95\% confidence.}
\end{table}

\subsection{X-ray bursts}
\label{sec:bursts}

\begin{figure*}[t]
  \includegraphics[width=\linewidth]{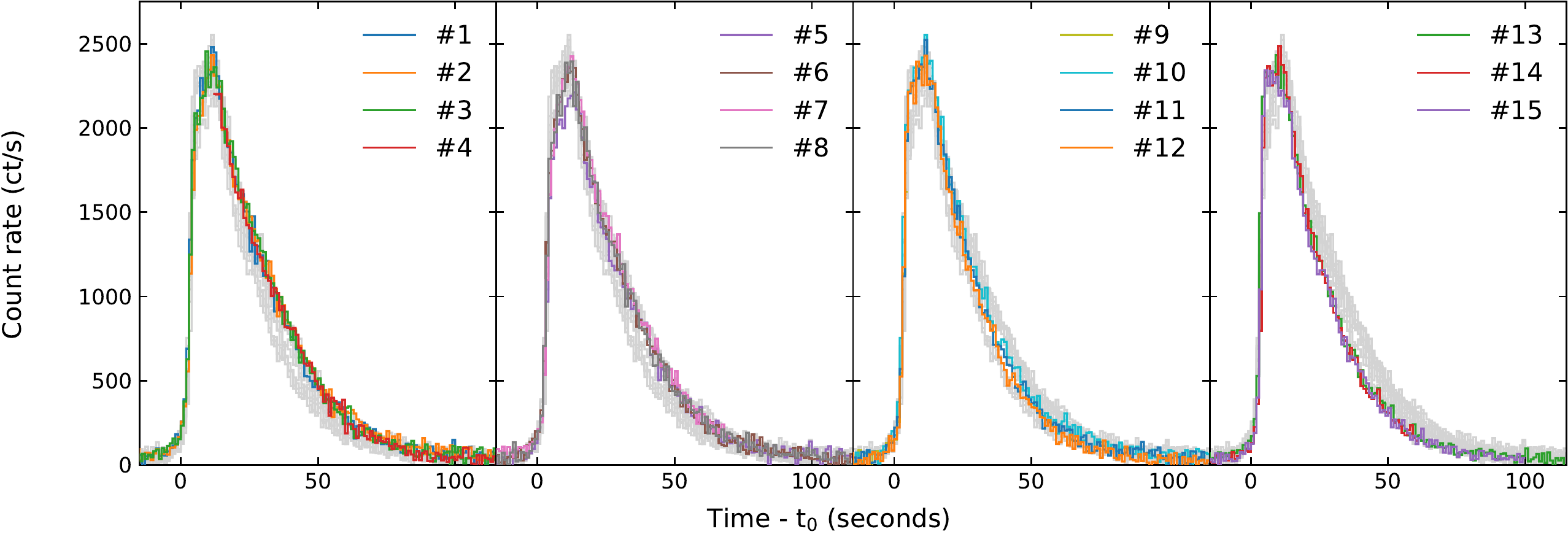}
  \caption{%
    Light curves of the 15 X-ray bursts observed from \src.  These light curves
    are calculated in the $0.5-10$ keV energy range at $1\s$ resolution, with
    the preburst count-rate subtracted. In each panel a different subset of the
    bursts are highlighted in color (with the burst number as in Table
    \ref{tab:burst spectroscopy}), while the remaining bursts are shown in gray
    for comparison.
  } 
  \label{fig:burst profiles}
\end{figure*}

We observed 15 type I X-ray bursts from \src, all with very similar profiles.
To characterize these profiles we manually determined the burst onset, $t_0$,
measured the rise time as the time from onset to the peak of the $1/8\s$
time resolution light curve, and measured the exponential decay
timescale between $[t_0+10\s, t_0+100\s]$ (see Table \ref{tab:burst
spectroscopy}).
The light curves of bursts $\#1$ through $\#8$ are almost identical, with
a rise time of about $8\s$ and an exponential decay timescale of about $23$\s.
As the outburst progressed the X-ray burst profiles showed a modest shift in
shape toward faster rise times and shorter decay times. This effect is
illustrated in Figure \ref{fig:burst profiles}.

We investigated the energetics of the X-ray bursts through a time-resolved
spectroscopic analysis. First we extracted a preburst spectrum from $[t_0-125\s,
t_0-25\s]$ (except for burst $\#4$, where we used the final $100\s$ of the
pointing). We then extracted time-resolved spectra from the X-ray bursts by
dynamically dividing the burst epochs into multiples of $0.1\s$, such that each
temporal bin contained at least $1500$ events.  

We modeled the X-ray burst spectra in the $0.8-10\kev$ energy range using
\textsc{xspec} version 12.12 \citep{Arnaud1996}. Subtracting the preburst
emission as background, we described each burst spectrum as an absorbed
blackbody (\texttt{tbabs * bbodyrad}). This model provided statistically
acceptable $\chi^2$ values throughout each X-ray burst. The average absorption
column density is $(2.36\pm0.06)\E{22}\persq{cm}$, while the typical peak
blackbody temperature and normalization are $1.90\pm0.03\kev$ and
$300\pm10\km^2$ at 10\kpc, respectively. None of the observed X-ray bursts
showed evidence for photospheric radius expansion or spectral lines. In Table
\ref{tab:burst spectroscopy} we list the bolometric blackbody fluence and peak
flux measured in each burst.

While a simple blackbody model is sufficient for individual bursts, we note
that the best-fit $\chi^2$ values show a systematic increase around the peak 
intensity of each X-ray burst. This suggests that the persistent emission may be modestly
affected by the burst flux. Indeed, if we fit the spectra at the peak of each
X-ray burst jointly with the blackbody temperature and normalization tied across all bursts,
then we find that an absorbed blackbody spectrum no longer
provides a sufficient description of the data. As an alternative model, we
adopt the so-called $f_a$ method \citep{Worpel2013}.  We generated
a background spectrum for each preburst spectrum using the \nicer 3C50
background model \citep{Remillard2022} and modeled each of the preburst
spectra using an absorbed disk blackbody plus a thermally Comptonized continuum
(\texttt{nthcomp}, \citealt{Zdziarski1996, Zycki1999}).  Finally, the X-ray
burst spectra were modeled using an absorbed blackbody plus the preburst model,
where the preburst component was multiplied with a variable factor, $f_a$. This
approach improved the $\chi^2/$dof from 985/863 to 945/862 (factor 10 improvement
in the p-value) with $f_a = 1.22\pm0.04$ at peak burst intensity. 

\begin{table*}[t]
  %\centering
  \movetableright=-1.00in
  \caption{%
    Detected X-ray bursts
    \label{tab:burst spectroscopy}
  }
  \newcommand{\flunit}{$\times10^{-7}$ erg/cm$^2$}
  \newcommand{\fxunit}{$\times10^{-8}$ erg/s/cm$^2$}
  %\begin{tabular}{l l l l l l l l l l}
  \begin{tabular}{c c c c h c c c c c c}
    \hline \hline
    ID  &  ObsID       & MJD            & Note & Peak rate       & Fluence           & Peak flux       &  Rise    & $\epsilon$     & $\tau$ & $\alpha$ \\ 
    ~   &  ~           & (TDB)          & ~    & (ct/s)          & (\flunit)         & (\fxunit)       &  (s)     & (s)            & (s)    & ~        \\ 
    \tableline
    1   &  5533010101  &  59738.793937  & ~    & 3360 $\pm$ 160  &  $ 9.5 \pm 0.7$  & $ 4.1 \pm 0.3$  &  $7.6$   & $23.2 \pm 0.2$  & $23.1$ &          \\
    2   &  5533010101  &  59738.976061  & SAA  & 3370 $\pm$ 160  &  $ 9.9 \pm 0.8$  & $ 4.6 \pm 0.4$  &  $8.8$   & $24.3 \pm 0.2$  & $21.7$ &          \\
    3   &  5533010102  &  59739.366594  & ~    & 3470 $\pm$ 170  &  $ 9.7 \pm 0.7$  & $ 4.2 \pm 0.4$  &  $7.4$   & $23.5 \pm 0.3$  & $23.1$ &          \\
    4   &  5533010103  &  59740.537525  & Tail & 3360 $\pm$ 160  &  $>6.7 \pm 0.5$  & $>4.0 \pm 0.4$  &  -       & $23.6 \pm 0.3$  &  -     &          \\
    5   &  5533010103  &  59740.790575  & ~    & 3240 $\pm$ 160  &  $ 9.1 \pm 0.7$  & $ 4.3 \pm 0.4$  &  $6.9$   & $23.5 \pm 0.3$  & $21.2$ &          \\
    6   &  5533010104  &  59741.633451  & ~    & 3520 $\pm$ 170  &  $ 9.4 \pm 0.7$  & $ 4.3 \pm 0.3$  &  $9.4$   & $22.7 \pm 0.2$  & $21.9$ &          \\
    7   &  5533010105  &  59742.413764  & SAA  & 3610 $\pm$ 170  &  $ 9.5 \pm 0.7$  & $ 4.3 \pm 0.3$  &  $7.9$   & $23.6 \pm 0.3$  & $22.3$ &          \\
    8   &  5533010105  &  59742.470702  & ~    & 3440 $\pm$ 170  &  $ 9.4 \pm 0.7$  & $ 4.2 \pm 0.3$  &  $7.3$   & $22.8 \pm 0.2$  & $22.4$ & $45\pm3$ \\
    9   &  5533010106  &  59743.103625  & Rise & 3010 $\pm$ 160  &  $>0.8 \pm 0.1$  & $>3.1 \pm 0.2$  & $>4.7$   & -               &  -     &          \\
    10  &  5533010801  &  59746.344881  & SAA  & 3340 $\pm$ 160  &  $ 9.3 \pm 0.6$  & $ 4.6 \pm 0.4$  &  $5.6$   & $21.4 \pm 0.2$  & $20.3$ &          \\
    11  &  5533010901  &  59747.308052  & ~    & 3340 $\pm$ 160  &  $ 9.0 \pm 0.6$  & $ 4.1 \pm 0.3$  &  $5.7$   & $20.6 \pm 0.2$  & $21.7$ &          \\
    12  &  5533010901  &  59747.567004  & ~    & 3440 $\pm$ 170  &  $ 8.8 \pm 0.6$  & $ 3.9 \pm 0.3$  &  $6.6$   & $19.4 \pm 0.3$  & $22.3$ &          \\
    13  &  5533011301  &  59751.757114  & ~    & 3080 $\pm$ 160  &  $ 8.4 \pm 0.6$  & $ 4.4 \pm 0.3$  &  $4.8$   & $19.5 \pm 0.2$  & $19.1$ &          \\
    14  &  5533011601  &  59754.160772  & ~    & 3232 $\pm$ 160  &  $ 8.3 \pm 0.5$  & $ 4.2 \pm 0.3$  &  $6.6$   & $18.6 \pm 0.2$  & $19.5$ & $58\pm4$ \\
    15  &  5533011601  &  59754.288382  & ~    & 2990 $\pm$ 150  &  $ 8.1 \pm 0.5$  & $ 4.0 \pm 0.3$  &  $4.8$   & $19.3 \pm 0.2$  & $20.2$ & $61\pm5$ \\
    \tableline
  \end{tabular}
  \flushleft
  \tablecomments{The MJD column lists the burst onset times. 
    Fluxes are bolometric. Columns $\epsilon$ and $\tau$ give the exponential decay timescale
    and the ratio burst fluence to peak flux, respectively. The $\alpha$ measurement of burst 14 follows
    from the $3\hr$ recurrence time obtained from simultaneous \nustar coverage (see text). All uncertainties
    give $1\sigma$ errors.}
\end{table*}

\subsubsection{Burst recurrence time}

Dividing the total unfiltered exposure (111\ks) by the number of detected X-ray
bursts, we estimate the average burst recurrence time at $2.1\hr$. This average
is close to the actual spacing observed between X-ray bursts: we find bursts
$\#7$ and $\#8$ are separated by $1.4\hr$, while bursts $\#14$ and $\#15$ are
separated by $3.1\hr$. 

\src was also observed with \nustar on 2022 June 23 \citep[MJD 	59753,][]{AtelChauhan22}. 
This \nustar observation contains four X-ray bursts, the first two of which
were not observed with NICER and the latter two being $\#14$ and $\#15$ in
our sample. This train of four X-ray bursts observed with
\nustar is consistent with a regular recurrence time of $3.0\hr$, while
a shorter recurrence time for these bursts ($1.5\hr$) is ruled out by
the joint \nicer and \nustar coverage. Hence,
these results suggest that \src is a regular burster with a recurrence time
that lengthens over the course of the outburst.

The $\alpha$ factor is defined as the ratio of the persistent fluence over the
burst fluence \citep{Galloway2020}, and can be estimated as \begin{equation} \alpha = \frac{F_{\rm
persist} \Delta t}{E_{\rm burst}}, \end{equation} where $F_{\rm persist}$ is
the bolometric flux of the persistent emission, $\Delta t$ the time between
successive X-ray bursts and $E_{\rm burst}$ the fluence of the burst.  We
estimate the $F_{\rm persist}$ by adding a \texttt{cflux} component to the
persistent spectra and measuring the flux between $0.01-100\kev$, while we
obtain the burst fluence from the time-resolved spectroscopy. For the three
X-ray bursts with a reliable measurement of the recurrence time ($\#8, \#14$,
and $\#15$) we find $\alpha$ values of 46, 58, and 61 (respectively).

\subsubsection{Burst oscillations}

To search the X-ray bursts for the presence of coherent burst oscillations, we
employed a sliding window search method. For each X-ray burst, we constructed
a light curve at $1/8192\s$ time resolution and applied a window to this
light curve of duration $T=2,4,8\s$. We then moved the window across the light curve
in steps of $T/4$. At each window position we calculated the power density
spectrum and searched for single bin powers between $528.6\pm5\hz$ that exceeded
the $3\sigma$ detection threshold calculated from the counting noise
distribution (correcting for the number of trials; the number of windows times
the number of powers per window). We applied this search strategy using all events 
in the $0.5-10\kev$ energy range.  No burst oscillation candidates were found. 

\clearpage
\section{Discussion}

We have presented the discovery of $528.6\hz$ pulsations from \src. Through
a timing analysis of the pulsations, we measured the binary ephemeris reported
in Table \ref{tab:timing solution}. From this ephemeris we find that the pulsar
mass function is $f_x = 4.8\E{-3}\msol$, implying a minimum companion mass of
$M_2>0.10\msol$ for a canonical $1.4\msol$ neutron star. 

\begin{figure}[t]
  \includegraphics[width=\linewidth]{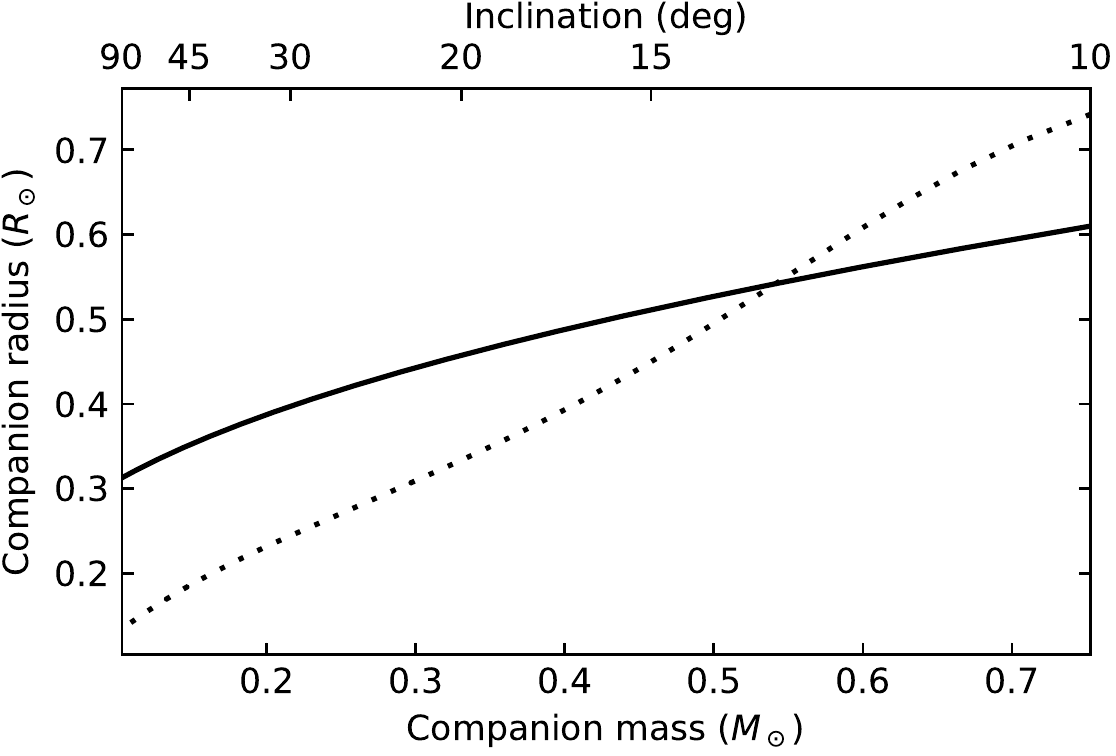}
  \caption{%
    Mass-radius relation of the companion star from the binary ephemeris (solid line)
    compared to the theoretical mass-radius relation of a zero-age main sequence
    star (dotted line).
  } 
  \label{fig:mass radius}
\end{figure}

If we assume that the companion star fills its Roche Lobe, we can use the Roche
lobe radius \citep{Eggleton1983} to calculate both the mass and radius of the
companion star as a function of the binary inclination.  This empirical mass-radius
relation intersects with the theoretical mass-radius relation for a zero-age
main-sequence star \citep{Tout1996} at a companion mass of $0.55\msol$ (Figure
\ref{fig:mass radius}). If the companion star is evolved or ablated by the
irradiation of the compact object, however, it will tend to a larger radius for
the same stellar mass, which means that we can treat this intersection as an
upper limit on the companion mass.  Indirectly, this mass limit then implies
a lower limit on the binary inclination of $13\arcdeg$.

We found that the pulse phase shows a complex evolution with time. The greater
part of the observed outburst, between $t=5-25$\,d, could be modeled using
a constant pulse frequency model. Outside this time range the pulse phase shows
systematic residuals relative to the model. 
Similar structural patterns in the phase residuals have been observed in
a number of other AMXPs \citep{Burderi2007, Hartman2008, Patruno2009b,
Bult2020a, Sanna2020}.
The common explanation for this behavior is that either the accretion torque is
measurably changing the stellar spin frequency over the course of the outburst,
or that the hotspot position on the stellar surface is not fixed, but shifts in
response to the changing accretion geometry. Of course these effects are not
exclusive, and could both be contributing factors.

For \src we were able to account for the phase residuals observed during the
outburst rise by employing a flux-bias model for the pulsar phase.
Interestingly, this model prefers a power law index of $\Gamma=1.2$,  which is
a shallower dependence than the $\Gamma\gtrsim2$ expected from a phase drift imposed through
a cumulative accretion torque \citep{Sanna2020}. Yet, this index is not
consistent with the hotspot position changes predicted from numerical
simulations of accreting pulsars either \citep{Kulkarni2013}. Hence, if the
hotspot is moving, its phase bias is not driven by the changing radius of the
magnetosphere, but instead nearly proportional to intensity and thus
the mass accretion rate. 

Even with the flux-bias model, we could not account for a discrete $0.3$ cycle
jump in the pulse phase observed in the final day of the outburst. Similar
late-time phase jumps have been observed from SAX J1808.4--3658 \citep{Burderi2006,
Hartman2008}, and have been attributed to a viewing geometry driven by the
receding accretion disk \citep{Ibragimov2009}. Given that the phase jump in
\src occurs right before the source drops below the detection level, it is
plausible that a similar mechanism may be at play here.

\subsection{Constraints from the X-ray bursts}

We found that all observed X-ray bursts showed a very similar evolution, both
in light curve and in their spectroscopy. The burst durations and $\alpha$
values point to ignition in a hydrogen rich environment \citep{Lewin1993,
Galloway2021}, indicating that the accreted material and thus the donor star
must be hydrogen rich. 
The peak fluxes are consistent within errors, with an average of
$(4.3\pm0.1)\E{-8}\fluxcgs$. Depending on the hydrogen abundance in the burst
fuel, the expected Eddington luminosity is $2.2-3.8\E{38}\lumcgs$
\citep{Kuulkers2003}, which yields an upper limit on the distance of
$d<6.5-8.6\kpc$.  The lower end of this range is associated with hydrogen rich
bursts, and is therefore preferred.

For three of the observed X-ray bursts we obtained a reliable measurement for
$\alpha$, finding that this factor increases from $46$ for burst $\#8$ to about
$60$ for bursts $\#14$ and $\#15$. Such evolution usually indicates a shift in
the average hydrogen abundance of the burst fuel \citep{Galloway2004}, which
would be consistent with the smaller fluence and decay timescale seen in later
bursts. Specifically, if we assume that all accreted matter burns during an X-ray
burst, then $\alpha$ follows from theory as \citep{Galloway2020} 
\begin{equation}
  \alpha = \frac{Q_{\rm gravity}}{Q_{\rm nuclear}} (1+z) \frac{\xi_{\rm burst}}{\xi_{\rm disk}},
\end{equation}
where $Q_{\rm gravity} = G M_{\rm ns} / R_{\rm ns}$ is the gravitational
binding energy, $Q_{\rm nuclear} = 1.35 + 6.05\overline X \mev\per{nucleon}$
the nuclear energy released by accreted matter \citep{Goodwin2019} (with
$\overline X$ the average hydrogen abundance in the burst fuel), $z$ the
gravitational redshift, and $\xi_{\rm burst}/\xi_{\rm disk}$ the ratio of the
burst and disk anisotropy \citep{Fujimoto1988}. If the CNO cycle
is stably burning hydrogen between the X-ray bursts then shorter recurrence
times give a higher $\overline X$ and thus a higher energy release per accreted
nucleon \citep{Galloway2021}. For solar abundances the observed increase
in recurrence time would then increase $\alpha$ by about 10\%, well short of
what is needed to explain the observations. This suggests that
some additional physical process is gradually changing the ignition
condition either as a function of time or mass accretion rate. Whether that
missing piece relates to an evolving accretion geometry, an inertia in the
burst train \citep{Woosley2004, Johnston2018}, a temperature evolution in the
neutron star crust \citep{Chenevez2010}, or perhaps some mixing effect related
to the ignition latitude \citep{Cavecchi2020} remains an open question.

\begin{acknowledgments}
\nolinenumbers
This work made use of data and software provided by the High Energy
Astrophysics Science Archive Research Center (HEASARC). 
PB acknowledges support from NASA through the NICER Guest Observer Program
and the CRESST II cooperative agreement (80GSFC21M0002).
SG acknowledges the support of the CNES.
DA acknowledges support from the Royal Society
GCM was partially supported by Proyecto de Investigaci\'on Plurianual (PIP) 0102 (Nacional de Investigaciones Cient\'{\i}ficas y T\'ecnicas (CONICET)) 
and from PICT-2017-2865 (Agencia Nacional de Promoci\'on Cient\'ifica y Tecnol\'ogica (ANPCyT)).
NICER science team members at NRL are supported by NASA under Interagency Agreement NNG200808A.
\end{acknowledgments}

\facilities{ADS, HEASARC, NICER}
\software{heasoft (v6.30), nicerdas (v9), xspec \citep{Arnaud1996}, tempo2 \citep{Hobbs2006}}

\end{document}